\documentclass[runningheads]{llncs}
\usepackage{color}
\usepackage{colortbl}
\RequirePackage{cite}
\usepackage{amssymb}
\usepackage{graphicx}
\usepackage{caption}
\usepackage{tcolorbox}
\usepackage{enumitem} \setlist{nosep}
\usepackage{adjustbox}
\RequirePackage{tikz,pgfplots}
\RequirePackage{algpseudocode}
\RequirePackage{algorithm}
\usepackage[utf8]{inputenc}
\usepackage{xcolor}
\usepackage[mathscr]{euscript}
\RequirePackage{amsmath,amssymb,bm}
\RequirePackage{enumitem}
\usepackage{multirow}
\usepackage{paralist}
\usepackage[textsize=scriptsize,color=yellow!25,linecolor=gray]{todonotes}
\usepackage{nameref}
\definecolor{darkgreen}{rgb}{0.0, 0.5, 0.0}

\begin{document}
\title{Joint Autoregressive and Graph Models\\
for Software and Developer Social Networks}
\titlerunning{Models for Software and Developer Social Networks}
%

\author{Rima Hazra\inst{1} \and
Hardik Aggarwal\inst{1} \and
Pawan Goyal\inst{1} \and
Animesh Mukherjee\inst{1} \and
Soumen Chakrabarti\inst{2} }
\authorrunning{Hazra et al.}
%

\institute{IIT Kharagpur, Kharagpur, India\\
\email{to\_rima@iitkgp.ac.in, hardik8464@gmail.com,}\\
\email{\{animeshm,pawang\}@cse.iitkgp.ac.in}\and
IIT Bombay, Mumbai, India\\
\email{soumen@cse.iitb.ac.in}}

\maketitle              
\begin{abstract}
Social network research has focused on hyperlink graphs, bibliographic citations, friend/follow patterns, influence spread, etc.  Large software repositories also form a highly valuable networked artifact, usually in the form of a collection of packages, their developers, dependencies among them, and bug reports.  This ``social network of code'' is rarely studied by social network researchers.  We introduce two new problems in this setting. These problems are well-motivated in the software engineering community but not closely studied by social network scientists. The first is to identify packages that are most likely to be troubled by bugs in the immediate future, thereby demanding the greatest attention.  The second is to recommend developers to packages for the next development cycle.  Simple autoregression can be applied to historical data for both problems, but we propose a novel method to integrate network-derived features and demonstrate that our method brings additional benefits. Apart from formalizing these problems and proposing new baseline approaches, we prepare and contribute a substantial dataset connecting multiple attributes built from the long-term history of 20 releases of Ubuntu, growing to over 25,000 packages with their dependency links, maintained by over 3,800 developers, with over 280k bug reports. 

\end{abstract}

\section{Introduction}
\label{sec:intro}


A rapidly growing, rich, complex and immensely valuable social network has garnered surprisingly little attention compared to the WWW hyperlink graph, follower-followee and retweet/repost/reply networks in social media platforms etc. This network is formed by software packages, the dependency graph that links them, their developers, and bug reports and discussions concerning them.  As a case in point, Linux, with its many flavors and adaptations, is a huge public software repository. It has tens of thousands of packages, connected by dependency and other links. Thousands of developers contribute to these packages, forming another aspect of the network.  In fact, the developer ecosystem evolves organically, rather than via central command-and-control chains. The network is highly dynamic, with accurately-maintained trace of evolution along with detailed logs of bug reports pertaining to different packages.  Business realities have made open-source software development viable even for commercial organizations, with notable examples like Tensorflow, ZFS, Ubuntu, Java, Postgres, etc.  A comparatively nascent and chaotic version of such self-organization of software networks can be found on github, gitlab and bitbucket.  

In this work we focus on the Ubuntu code repository.
Ubuntu, a Linux based distribution is a collection of many open source software/packages. The project encourages the community to contribute to the development and maintenance of one or more packages. For every package, there is a set of developers (often one) who are responsible for the maintenance of the package and keep track of all the changes to the package in a changelog\footnote{http://changelogs.ubuntu.com/changelogs/}, recording the sequence of bug fixes or other updates related to the package.

Unlike traditional social network tasks of centrality/prestige computation, influence or cascade prediction, the social network of software comes with novel tasks having strong motivation and relevance in the software management community.
\noindent{\bf Bug urgency ranking:}
The task is to rank packages that are likely to be most afflicted by bugs in the immediate future. Since there is no central command, the developer community has to autonomously discover the trouble spots. 
\noindent{\bf Developer recommendation:} For each package, the task is to propose the developers best suited to contribute in the immediate future. Compared to software corporations with top-down management, the developer community shows high levels of churn, making such prediction difficult.  
We know of no widely used public domain tools for predicting bug urgency or recommending developers for a given package.
While there are several articles on developer/commenter recommendation~\cite{Xuan:2015} in various community question answering sites, to our knowledge, none of them attempt to build a model to recommend the developers in software development platforms like Ubuntu.

\noindent\textbf{Our contributions and results ---} 

\noindent\textbf{{\em A new dataset}}: We contribute a substantial new 
dataset\footnote{http://doi.org/10.5281/zenodo.4092623} connecting multiple  software and developer artifacts built from the long-term history of 20 releases of Ubuntu, growing to over 25k packages with their dependency links, maintained by over 3800 developers, with over 280k bug reports. There are 25k unique nodes (packages) and 120k dependency links among these packages across the Ubuntu releases. 

\noindent\textbf{{\em Algorithms for bug urgency ranking}}: We propose autoregressive baselines that predict future bug urgency as a regression based on recent history, then augment them with novel ways to incorporate inter-package dependency graph signals, which result in enhanced ranking accuracy. We are able to achieve high rank correlation between gold and system rankings, for both the autoregressive and autoregressive+dependency models. For the most recent distribution (i.e., Zesty) in our data set, Spearman's rank correlation $\rho@25$ and Kendall's $\tau@25$ values are respectively 0.582 and 0.451 using only the autoregressive features. Inclusion of dependency features further improves both the correlation values ($\rho@25=0.60$ and $\tau@25=0.466$). If one considers the full rank list then we obtain $\rho=0.35$, $\tau=0.33$ for the autoregressive case and $\rho=0.38$, $\tau=0.35$ in case of autoregressive+dependency. For the full rank list the differences in the results between the autoregressive and autoregressive+dependency schemes are statistically significant ($p<0.01$ for both $\rho$ and $\tau$ as per Mann-Whitney U test~\cite{mannwhitney:1947}).

\noindent\textbf{{\em Algorithms for developer recommendation}}: In its most basic form, recommending developers for a package may be modeled as predicting a set given a sequence of past sets~\cite{seqOfsets:2018}. However, our data set has richer signals in both space (i.e., graph structure) and time, as well as features from bug reports, bug fix changelogs, etc.  Even a simple autoregressive approach is able to take advantage of these features and outperform baselines. For the most recent distribution, the Mean Reciprocal Rank (MRR) for the autoregressive approach is ${\sim}0.788$ as compared to $0.772$ for the best performing baseline. Additional benefits are also obtained from the dependency relations (MRR ${\sim}0.793$).  Subject to some reasonable assumptions, we also compute {upper bounds} for autorgressive and autoregressive+dependency schemes as $0.8096$ and $0.8445$ respectively, which gives ample scope of improvement in future.

\section{Related work}
\label{sec:related_work}

Recommendation systems are nowadays becoming available to assist developers in various activities --- from reusing code \cite{janjic2014reuse} to writing effective bug reports~\cite{naguib2013bug,anvik2006automating}.
  
\noindent\textbf{{\em Developer recommendation approaches}}: We witness a growing volume of literature on developer recommendation for crowdsourced tasks. Mao et al.~\cite{mao2015developer} employed content-based recommendation techniques to automatically match tasks and developers for the TopCoder platform. Related work~\cite{10.1145/2961111.2962594} recorded a task-quitting rate of 82.9\% among TopCoder developers.  Ye et al.~\cite{7557978} proposed four problems that limit the effectiveness of existing methods at recommending suitable developers. Tunio et al.~\cite{8026018} studied the impact of personality on task selection in crowdsourcing software development. 

\noindent\textbf{{\em Package dependency networks}}: De Sausa et al.~\cite{de2009analysis} presented an analysis of the package dependency on Debian GNU/Linux.  Kikas et al.~\cite{kikas2017structure} studied the structure and evolution of package dependency networks of JavaScript, Ruby, and Rust ecosystems. Decan et al.~\cite{decan2016topology} showed that experimental results related to software packages belonging to a single software ecosystem fail to generalise to other ecosystems because of the diversity of their structure.

We know of no widely used approach that uses package dependency networks for developer recommendation. Also, we find limited research on Dirichlet based sampling approach in recommendation and ranking. In this paper, we combine  the two paradigms for the two tasks that we solve --- bug urgency prediction and developer recommendation. 

\section{Dataset}
\label{sec:datasets}

Ubuntu is a free and open-source Linux distribution based on Debian, released for Desktop, Server and IoT deployment\footnote{\protect\url{https://www.ubuntu.com/\#download}}. 
It is released every six months, with long-term support (LTS) releases every two years. 
Our data consists of three parts: (i)~Ubuntu packages and the dependencies among these packages,  (ii)~developers of Ubuntu packages responsible for bug fixes and other updates and the maintenance of change logs, and (iii)~bug(s) associated with each package.  For our experiments we only use 20 non-LTS versions (binary-amd64) published between April 2004 and April 2017.  Dataset details follow in this section.

\noindent\textbf{{\em Ubuntu packages and their dependencies}}: Each Ubuntu distribution contains a collection of binary packages. Binary packages are made for different types of architectures like AMD64, i386 etc. For each distribution, we collected binary packages and their dependencies. The most prevalent form of dependency between a pair of binary packages is referred to as \textit{depends}\footnote{There are other types of relations also in the dataset like \textit{recommends}, \textit{suggests} and \textit{conflicts} which are very infrequent.}. A binary package $P_{i}$ \textit{depends} on another binary package $P_{j}$ if $P_{j}$ is required to build and install $P_{i}$. ``Dependee'' denotes a binary package ($P_j$) on which another binary package ($P_i$) depends. In the rest of this paper the dependency network that we refer to is built from this \textit{depends} relation.

\begin{figure}[h]
\centering\scriptsize
\begin{minipage}{.5\textwidth}
\resizebox{1\textwidth}{!}{
\begin{tikzpicture}
  \node (s) [draw=red, fill=gray!30] {systemd};
  \node (s1) [below=15mm of s, draw=red, fill=gray!30] {glibc};
  \node (s2) [left=1cm of s1,  draw=red, fill=gray!30] {libseccomp};
  \node (s3) [right=1cm of s1,  draw=red, fill=gray!30] {iptables};
  \node (d2) [above right=1cm of s, draw=red, fill=red!10] {Martin Pitt};
  \node (d3) [above left=1cm of s,draw=red, fill=red!10] {Dimitri John Ledkov};
  \node (d4) [below=1cm of s1,draw=red, fill=red!10] {Adam Conrad};
  \draw[->] (s) edge (s1);
  \draw[->] (s) edge (s2);
  \draw[->] (s) edge (s3);
  \draw[->] (s) edge [draw, dashed, thick] (d2);
  \draw[->] (s) edge [draw, dashed, thick] (d3);
  \draw[->] (s3) edge [draw, dashed, thick] (d2);
  \draw[->] (s2) edge [draw, dashed, thick] (d3);
  \draw[->] (s1) edge [draw, dashed, thick] (d4);
\end{tikzpicture}
}
\caption{Dependees and developers of the `systemd' package. Gray nodes: software packages, Red nodes: developers. Solid and dotted lines represent \textit{`source dependency'} and \textit{`contributed by'} relationships, respectively. `Dimitri John Ledkov' is a developer associated with both `systemd' and its dependee `libseccomp'.}
\label{fig:example_graph}
\end{minipage}
\begin{minipage}{.45\textwidth}
\centering
\includegraphics[width = 0.8\textwidth, height=4.5cm]{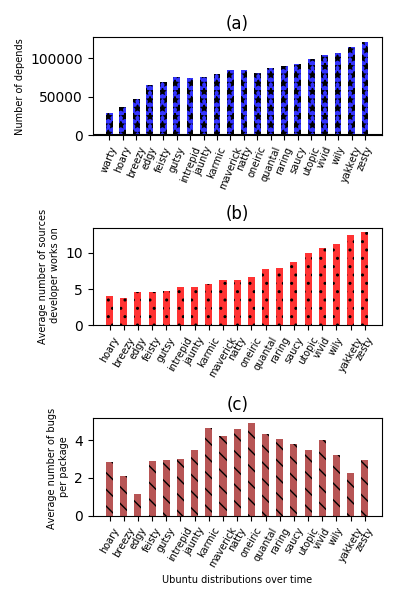}
\caption{\label{fig:depn_src_bugs}(a)~The no. of ``depends'' dependencies in each Ubuntu distribution. (b)~The average no. of source packages the developers worked on across the distributions. (c)~The average no. of bugs per source package (with non-zero bugs).} 
\end{minipage}
\end{figure}

A source package, on building, may generate a set of binary packages\footnote{\protect\url{https://askubuntu.com/questions/357295/what-is-difference-between-binary-and-source-file}}.  E.g., the source package ``0ad''\footnote{https://packages.ubuntu.com/source/xenial/0ad} contains the binary packages, ``0ad'' and ``0ad-dbg''.  We consider source packages and their dependencies in our experiments. We chose source packages instead of binaries since the source packages have a unique identify with source codes unlike binary packages which may correspond to compiled codes from different architectures.
We present all our results for the three most recent distributions -- `Wily', `Yakkety' and `Zesty' which have 22799, 24609 and 25648 source packages respectively. An example dependency graph is shown in Figure~\ref{fig:example_graph}. The package `systemd' and its dependees `libseccomp', `glibc' and `iptables' are shown in gray in the figure. 

The number of ``depends'' package dependencies across the Ubuntu distributions are reported in Figure~\ref{fig:depn_src_bugs} (a) (increases as time progresses).

\noindent\textbf{{\em Developers of Ubuntu packages}}: The `changelog'\footnote{\raggedright E.g.,~\protect\url{http://changelogs.ubuntu.com/changelogs/pool/universe/0/0ad/0ad\_0.0.20-1/changelog}} of a source package contains add/remove/update information about the codebase and bugs (resolved bugs) associated with the package. It also contains the urgency level, name of the developer and timestamp of that \textit{change}.  Packages evolve at diverse paces, and distributions take snapshots at discrete points in time. We have collected changelogs of source packages and mapped them to Ubuntu distributions. These change logs allow us to associate every developer with one or more packages for each distribution. An example of the developer-package relation is illustrated in Figure~\ref{fig:example_graph}. In our dataset, we observed that a particular source typically (but not always) has a single developer in  each distribution. Over time the average number of packages a developer contributes to is reported in Figure~\ref{fig:depn_src_bugs}(b) (shows an increasing trend). 

\noindent\textbf{{\em Bugs associated with Ubuntu packages}}: Ubuntu releases do not provide a straightforward way to recover the bugs (along with their meta data) that are associated with a particular distribution. We therefore collected 280k bugs along with all available information from Launchpad.\footnote{\protect\url{https://launchpad.net/}}  We next associated a bug with a particular distribution if that bug had been created within six months from the release of that distribution. Next we mapped these bugs associated with a distribution to the corresponding source packages. The number of bugs per package, averaged over all packages and all distributions, is $\sim3.4$ (considering all packages that have non-zero reported bugs). The average number of bugs per package over time is reported in Figure~\ref{fig:depn_src_bugs}(c). The plot shows that in early versions of Ubuntu, fewer bugs were reported, followed by a sharp increase and then a final decline. This possibly indicates that as the software became more complex and popular, the number of bugs reported grew quickly.  However, it settled down at later time point due to the consolidated rectification efforts made by developers.

\section{Notation and preliminaries}
\label{sec:Notation}

We have a set of $T$ \textbf{distributions} of a large software system (such as Ubuntu) indexed as $t\in[T]=\{1,\ldots,T\}$, where $t$ represents discrete, ordinally comparable time and equivalently, distributions. An example is $\text{vivid} < \text{wily}$ where vivid and wily are the Ubuntu distributions.
There is a universe of $S$ \textbf{packages} indexed by $s\in [S]=\{1,\ldots,S\}$.  An example is $s = \text{glibc}$.
(By `packages', we will mean ``source packages'' in the context of Ubuntu.  A source package may build to multiple binary packages, but developers are naturally assigned to source, not binary packages.)
Not all packages $s$ may be present in all distributions~$t$. Packages can be removed and later restored. For each package $s$, there is a package size $\text{ps}(s,t)$ which is the sum of the sizes of its binaries at distribution $t$.
There is a universe of $D$ \textbf{developers}, indexed as $d\in[D]=\{1,\ldots,D\}$.  An example is $d = \text{torvalds@linux.org}$.
A developer may contribute to many packages at various time steps (distributions).
Let $\text{devs}(s,t)\subset[D]$ denote the set of developers associated with package $s$ at time~$t$.
There is a universe of $B$ bug records, indexed as $b\in[B]=\{1,\ldots,B\}$.  One bug record attaches to a single package at a single distribution.
Let $\text{bugs}(s,t)\subset[B]$ denote the set of bugs associated with package $s$ at time~$t$. We denote the heterogeneous graph constructed at each time step $t$ as~$G_t$.  $G_t$ comprises two types of nodes --- source packages $s$ and developers $d$. Edges $s\to s'$ represent \textit{`source dependency'} relationships. The term ``dependent'' corresponds to the source package ($s$) that depends on another source package. ``Dependee'' represents the source package ($s'$) on which the dependent depends. Edges $s\to d$  represent \textit{`contributed by'} relationships.
Figure~\ref{fig:example_graph} shows an illustrative graph fragment at $t = \text{zesty}$ for a source package `systemd'. Note that a developer can work on multiple packages at the same time. For example, `Dimitri John Ledkov' is developer of both `systemd' and its dependee package `libseccomp'. 
Let the set of in-neighbors and out-neighbors of the target source package $s$ at time point $t$ be $S_{\text{IN},t}$ and $S_{\text{OUT},t}$ respectively.

\section{Bug urgency ranking}
\label{sec:BugUrgency}

Suppose we observe the evolution of the software ecosystem from time step $1$ through $t-1$.  In other words, we observe $G_\tau$ for $\tau\in[1,t-1]$ along with developer and bug sets associated with each package and time step.  Now, for time step $t$, our goal is to predict $|\text{bugs}(s,t)|$ for all packages~$s$.  
More practically, we want to sort packages $s$ in decreasing order of $|\text{bugs}(s,t)|$ and report the top ranks to attract the attention of central members of the developer community, so that they can solicit and allocate more programmer resources.  With that motive in mind, we are generally interested in predicting only the \emph{relative bug density} among packages in the next release.  While evaluating, we naturally have access to the gold $\text{bugs}(s,t)$, and we can therefore compare the system and gold rankings using various rank correlation measures.

\noindent\textbf{{\em Pure autoregressive approach}}: In this approach, we attempt to estimate the rank of source packages based on the bugs reported at earlier time points.  From the bugs information, we computed the number of bugs ($|\text{bugs}(s,t)|$) of each source package ($s$) for each time point $t$.  Specifically, we extract autoregressive features from earlier time points and predict the $|\text{bugs}(s,t)|$ for the current time point. For each source package, we consider two autoregressive features from the previous two time points, i.e., $|\text{bugs}(s,t-1)|$, $|\text{bugs}(s,t-2)|$, $\text{ps}(s,t-1)$ and $\text{ps}(s,t)$. 

We also tried to use the same features from even earlier time points. However, their contribution to the overall prediction performance is negligible compared to the last two time points and hence they are ignored. We observe that the bug history of two previous time points always contributes more than the package sizes in the prediction.  Our intuition behind utilizing package size is that, if the package size changes from the last time point to the current time point, then the package should contain some new updates. For example, for the package ``systemd'' in the ``Zesty'' distribution, the package size and the number of bugs are 6.12 MB and 21 respectively. In the previous distribution ``Yakkety'', the package size and number of bugs were 4.43 MB and 15 respectively. This and other similar observations made us hypothesize that the package size at time point $t$ might have potential correlation with the number of bugs at~$t$. 

\noindent\textbf{{\em Inclusion of network features}}: We hypothesize that the bugs in a particular source package could potentially induce bugs in its dependees as well as dependents. For instance, in distribution ``Zesty'', the ``systemd'' package has 21 bugs whereas in the immediate previous distribution (``yakkety'') this number is 15. The observed rise may be attributed to the very large number of bugs (60) associated with one of the in-neighbours (``linux'') of ``systemd'' in the previous distribution ``Yakkety''. Overall, across our full dataset, the Pearson's correlation between the bugs of a source package at time point $t$ and the bugs of its in-neighbors/out-neighbors at the previous time point $t-1$ lies between approximately $0.18$ and $0.28$. This makes us further confident that positive benefits could be obtained by considering the previous time point bugs of in-neighbours and out-neighbours as additional features.

Therefore, along with the autoregressive features, we also use the dependency features, i.e., the number of bugs of the in-neighbors and the out-neighbors. We deduce four such features detailed below. 

\noindent\textit{In-neighbor bugs}: We use the bugs of the in-neighbor source packages of $s$ from the previous time point as features. In particular, we consider the following two features: $\max_{s' \in S_{\text{IN},t}}(|\text{bugs}(s',t-1)|)$ and $\text{median}_{s' \in S_{\text{IN},t}}(|\text{bugs}(s',t-1)|)$ which are respectively the maximum and the median bug counts of the in-neighbours of the package $s$ from the previous time point ($t-1$). 

\noindent\textit{Out-neighbor bugs}: Similarly, as above, we use the bugs of the out-neighbor source packages of $s$ from the previous time point as features. Here we consider $\max_{s' \in S_{\text{OUT},t}}(|\text{bugs}(s',t-1)|)$ and $\text{median}_{s' \in S_{\text{OUT},t}}(|\text{bugs}(s',t-1)|)$ which are respectively the maximum and the median bug counts of the out-neighbours of the package $s$ from the previous time point ($t-1$). 

Note that in this case we predict $|\text{bugs}(s,t)|$ using both sets of features above as well as the autoregressive features, and, thereby, rank the source packages.

\section{Developer recommendation}
\label{sec:authors_feature}
Suppose we observe the evolution of the software ecosystem from time step $1$ through $t-1$. In other words, we observe
$G_\tau$ for $\tau\in[1,t-1]$ along with developer and bug sets associated with each package and time step.
Now, for time step $t$, our goal is to predict $D_{s,t}$.  This time, we are interested in \emph{ranking} developers by decreasing suitability for $(s,t)$. Suppose the system returns a ranked order $R_{s,t}$ over a suitable subset of developers. From the gold developer set $D_{s,t}$, we know the `relevant' or `good' positions, and can use any ranking evaluation measure such as MRR.

For this experiment, we consider two polices for creating candidate set of developers for $(s,t)$. (1)~main list: This list contains the developers who worked on the same source package $s$ in the previous distributions. (2)~also use the dependency list, which contains developers who worked on the neighbors (in-neighbors, out-neighbors) of the source package $s$ in the previous distributions. While the first policy goes well with the autoregressive features, the second policy is used while making use of dependency graph.

\noindent\textbf{{\em Model architecture and inference}}: Our objective is to rank a set of candidate developers for each source package and assign the top ranked developer in the test distro for that source package. Let us fix a source package $s$. $D_{s,\le t}$ is developer set for package $s$ up to time $t$. The developers could be collected from $s$'s history only or accessed via network.``$\le t$'' may mean $[t-K, t]$ depending on sliding window width $K$. Next we train a globally shared model $\theta$ for each such horizon $h$ (see Algorithm~\ref{algo:algo1}) 
We observe $D_{s,<h}$ for each package $s$. Next, we predict $D_{s,h}$, incur any loss and update $\theta$. Model $\theta$ induces a score on every developer $d\in D_{s,<h}$. For simplicity call this score $\theta(d)$. For all $d_+ \in D_{s,h}, d_- \not\in D_{s,h}$, we want $\theta(d_+) \gg \theta(d_-)$. In our evaluation protocol, all gold developer assignment at time $T$ are used as instances. For evaluating a system at time $T$ alone, apply model $\theta$ 
on candidate set $D_{s,< T}$ (note, not $T$) and predict ranking $R_{s,T}$ (meaning, sort by decreasing $\theta(d)$) which is evaluated wrt $D_{s,T}$. We categorize the developers present in candidate set in two clusters (i)~positive developers, (ii)~negative developers. Positive developers are the developers who are present in $D_{s,<h}$ as well as in $D_{s,h}$. Negative developers are the developers who are present in $D_{s,<h}$ but may leave for other reasons in $D_{s,h}$. Let $\bm{x}_{s,d+}$ denote feature vectors representing developers in the positive developer set. Similarly let $\bm{x}_{s,d-}$ denote feature vectors representing developers in the negative developer set. We outline a top level overview of the model architecture in Algorithm~\ref{algo:algo1}.

\begin{algorithm}[h]
\scriptsize
\caption{\label{algo:algo1} Top-level model architecture for developer recommendation.}
\begin{algorithmic}
\State initialize $\theta$
\State prepare batch loss expression (see below)
\For{horizon $h= T-K,\ldots,T-1$}
\For{each package $s$}
\State collect $D_{s,<h}$
\State two policies: either same package or via network;
\State positive devs $D^+_{s,h}$ are $D_{s,<h}\cap D_{s,h}$
\State negative devs $D^-_{s,h}$ are $D_{s,<h}\setminus D_{s,h}$
\If{$D^+_{s,h} \ne \varnothing$ and $D^-_{s,h}\ne\varnothing$}
\State represent each developer $d$ wrt package $s$ as $\bm{x}_{s,d}$, 
\State ``an instance'' $\bigl\langle (s,h); \{ \bm{x}_{s,d}:d \in D^+_{s,h} \},
\{ \bm{x}_{s,d}: d\in D^-_{s,h} \} \bigr\rangle$
\State batch loss has been drawn depending on the model chosen (LR/MLP)
\State call SGD optimizer for one batch to update~$\theta$
\EndIf
\EndFor
\EndFor

\State trained model $\theta$ available at this point
\For{each package $s$}
\State collect $D_{s,< T}$
\State prepare feature vectors $\bm{x}_{s,d}$ for each
$d \in D_{s,< T}$ and apply $\theta(\bm{x}_{s,d})$
\State sort candidate~$d$s by decreasing score
\State evaluate ranking $R_{s,T}$ wrt gold $D_{s,T}$
\EndFor
\end{algorithmic}
\end{algorithm}

We employ two different models -- (i)~Logistic Regression (LR) and (ii)~Multilayer Perceptron (MLP) to estimate $\theta$.

\noindent \textit{LR model}: Our optimisation function is $\theta(\bm{x}_{s,d}) = \sigma(\text{matmul}(\bm{x}_{s,d}, W) + b)$ and the loss expression is $loss = \max(0, (\theta(\bm{x}_{s,d_-}) - \theta(\bm{x}_{s,d_+}) + 1))$. Here $W$ and $b$ are the learnable parameters that we fit using stochastic gradient descent.

\noindent \textit{MLP model}: We use a feedforward neural network with one hidden layer. The model equations are $layer1(\bm{x}_{s,d}) = \tanh(\text{matmul}(\bm{x}_{s,d}, W_1) + b_1)$ and $\theta(\bm{x}_{s,d}) = \text{matmul}(layer1(\bm{x}_{s,d}), W_2)+ b_2$ respectively. The loss function is $loss = cost + L_2~\text{penalty}$, where the $cost = \sigma(\text{multiply}(a, (\theta(\bm{x}_{s,d_-}) -\theta(\bm{x}_{s,d_+}) - b)))$ and $a = \log(1 + \exp(\alpha)), b = \log(1 + \exp(\beta))$. Thus we maintain $a,b>0$ while $\alpha$ and $\beta$ are unconstrainted. 
$W_1$, $b_1$, $W_2$, $b_2$, $\alpha$ and $\beta$ are the learnable parameters. The $L_2$ penalty is calculated over all the learnable parameters. We use stochastic gradient descent. 

\noindent\textbf{{\em Feature construction}}: Next we discuss how to compute the features $\bm{x}_{s,d}$.

\noindent\textit{Pure autoregressive features}: In this approach, each developer $d$ for a source package $s$ at time $t$ is scored based on autoregressive features. From the changelog, we compute four features --- \textit{number of} \textit{high}, \textit{medium} and \textit{low urgency level} of packages on which the developer has worked, and the \textit{number of bugs closed by the developer}. In addition, we introduce a feature which captures the recency --- that is, whether the candidate developer worked on this package at time $t-1$.

\noindent\textit{Inclusion of network features}: Once again, like bug urgency prediction, we leverage  dependency links to improve developer recommendation. Our hypothesis is that developers who have recently contributed to one or more of the in(out)-neighbour packages of a source package should have a greater chance of contributing to the source package itself. This is because, the developers naturally acquire parts of the necessary skill set to contribute to the source package by having already contributed to its closely related packages (in- and out-neighbours) in the recent past. Thus, in addition to the autoregressive features, we add a set of dependency features from previous $K$ distributions -- $(t-1)$, $(t-2)$, $(t-3)$ and so on up to $(s,t-K)$. The features are \textbf{(i)} $K-1$ binary features telling whether the candidate developer was present in main developer list of $(s,t-i)$ where $i \in [2, K]$, \textbf{(ii)} $K$ binary features telling whether the candidate developer was present in the neighbor list of $(s,t-i)$ candidate distribution where $i \in [1, K]$, \textbf{(iii)} if the candidate developer is present in the main list of $(s,t-1)$ as well as in at least one of the neighbor list of $(s,t-2)$, $(s,t-3)$, and so on up to $(s,t-K)$, \textbf{(iv)} if the candidate developer is present in the neighbor list of $(s, t-1)$ as well as in at least one of the main list of  $(s,t-2)$, $(s,t-3)$, and so on up to $(s,t-K)$, and \textbf{(v)} if the candidate developer is present in the neighbor list of $(s, t-1)$ as well as in at least one of the neighbor list of  $(s,t-2)$, $(s,t-3)$, and so on up to $(s,t-K)$.

\section{Experiments and results}
\label{sec:exp}
\subsection{Bug urgency ranking}
\noindent{\textbf{\em Experimental setup}}: For this experiment, we consider only those source packages whose bug count in any of previous 10 distributions is non zero. We use a train-test split of 5:1 to train and evaluate our model. Let us say we have to predict the bug urgency of all the source packages at time point $t$. In order to train the model we use the data for all the source packages that appear in the $K$ previous time points. For each time point $(t-1)$ to $(t-K)$ and for every source package $s$ we calculate the autoregressive and dependency features as discussed above; accordingly, the training label for each time point is $|\text{bugs}(s,\cdot)|$ where the $\cdot$ ranges from $(t-1)$ to $(t-K)$. To train the model, we use the random forest regressor\footnote{One may argue that more complex models like point processes could be a possible choice. However note that we only have 20 time points and therefore such complex models cannot be trained sufficiently.}. We choose hyperparameters from the following intervals -- n\_estimators: $[100, 900]$, max\_depth: [4,7],  min\_samples\_split: $[4, 28]$ , min\_samples\_leaf: $[20,80]$, random\_state: $[0,8]$ . We used grid search to find the best parameter combination for both the autoregressive and the dependency approaches.

\noindent{\bf {\em Evaluation}}: For a given time point $t$, we rank the source packages based on ground truth $|\text{bugs}(s,t)|$ and the predicted $|\text{bugs}(s,t)|$. We use average ranking method to rank both the score lists. We use Spearman's rank correlation $\rho$ and Kendall's $\tau$ for evaluation. We report $\rho@25$, $\tau@25$ and the $\rho$, $\tau$ for the (quite large) full rank list\footnote{The full rank list has 4K packages on average.} (see Table~\ref{tab:bug_ranking_spearman}). We observe that for the most recent time point (i.e., Zesty) the the correlation values are pretty decent ($\rho@25=0.582$, $\tau@25=0.451$). Use of dependency features bring further benefits ($\rho@25=0.60$, $\tau@25=0.466$). In fact, for the full rank list also the results using the autoregressive+dependency features are quite good and are significantly different ($p<0.01$, Mann-Whitney U test) from those using only autoregressive features.

\begin{table*}[t]
\centering
\scalebox{0.78}{
\begin{tabular}{c|c|c|c|c} \hline
\multirow{2}{*}{\bf Distribution} & {\bf $\rho$@25} & {\bf $\tau$@25} & {\bf $\rho$} & {\bf $\tau$} \\ 
\cline{2-5}
&{\bf (auto, +depn)} & {\bf (auto, +depn)} & {\bf (auto, +depn)} & {\bf (auto, +depn)}  \\ \hline
\textbf{Wily} Werewolf & (0.546, 0.546) & (0.407,  0.407) & \cellcolor{green!20}(0.447, 0.454)**  & \cellcolor{green!20}(0.367, 0.371)** \\ \hline
\textbf{Yakkety} Yak & \cellcolor{green!20}(0.488, 0.498)& (0.331, 0.331) & \cellcolor{green!20}(0.260, 0.276)** & \cellcolor{green!20}(0.218, 0.240)**  \\ \hline 
\textbf{Zesty} Zapus & \cellcolor{green!20}(0.582, 0.603) & \cellcolor{green!20}(0.451, 0.466) & \cellcolor{green!20}(0.354, 0.380)** & \cellcolor{green!20}(0.328, 0.351)** \\ \hline 
\end{tabular}}
\caption{Spearman's $\rho$ and Kendall's $\tau$ for bug urgency ranking --- autoregressive only (auto), autoregressive + dependency (+depn). \textcolor{darkgreen}{Green} cells indicate cases where dependency features bring in additional benefits. ** indicates that the values of $\rho$ and $\tau$ for (auto) and (auto, +depn) are significantly different ($p<0.01$ as per Mann-Whitney U test).}\label{tab:bug_ranking_spearman}
\end{table*}

\begin{table*}
\scalebox{0.76}{
\begin{tabular}{|c|c|c|c|c|c|c|c|} \hline

\multirow{2}{*}{\bf Distribution}&\multicolumn{4}{c|}{Autoregressive} & \multicolumn{3}{c|}{Autoregressive + dependency}\\ 
& \multicolumn{4}{c|}{(auto)} & \multicolumn{3}{c|}{(auto+depn)}\\ 
\cline{2-8}
 & {\bf Our model} & \textbf{Majority} & {\bf SeqOfSets} & \textbf{Upper Bound} & {\bf Our model}  & \textbf{Majority} & \textbf{Upper Bound} \\ \hline
 \textbf{Wily} Werewolf & \cellcolor{blue!20}0.748** & 0.736  & \cellcolor{red!20}0.703 & \cellcolor{yellow!20}0.768  & \cellcolor{green!20}0.763$^\text{++}$ & 0.753  & \cellcolor{yellow!20}0.844\\
\cline{1-8}
 \textbf{Yakkety} Yak & \cellcolor{blue!20}0.628**  & 0.607 & \cellcolor{red!20}0.592 & \cellcolor{yellow!20}0.660  & \cellcolor{green!20}0.642$^\text{++}$ & 0.631  & \cellcolor{yellow!20}0.740\\
\cline{1-8}
\textbf{Zesty} Zapus & \cellcolor{blue!20}0.788**  & 0.773 & \cellcolor{red!20}0.725 & \cellcolor{yellow!20}0.810 & \cellcolor{green!20}0.794 & 0.785  & \cellcolor{yellow!20}0.844\\ \hline
\end{tabular}}
\caption{Developer recommendation: MRR values comparing our method with different baselines. **: Our results are significantly different from both baselines ($p<0.001$ for sequence of sets, $p<0.05$ for majority, Mann-Whitney U test). $^\text{++}$: Our results are significantly different from majority baseline ($p<0.01$, Mann-Whitney U test).}
\label{tab:WeVsSeqOfSets}
\end{table*}

\subsection{Developer recommendation}

\noindent{\bf {\em Upper bound}}: We first compute an achievable upper bound using the two policies for creating candidate set as discussed earlier i.e., (i)~main list and (ii)~main list + dependency network. If the developer of a source package at test distro is present in the candidate developer set then the rank of the developer is set to 1.\\

\noindent{\bf  Baselines ---}\\
\noindent{\bf {\em Sequence of Sets}}: In~\cite{seqOfsets:2018}, the authors proposed a stochastic model to capture the sequential behaviour of different tasks (such as sending emails, academic collaboration etc.). They proposed two parameters --- (i)~a correlation parameter (ii)~a vector of recency parameters. The correlation parameter measures the chance of repeating the earlier set in future. The recency parameters measure the similarity of a set with the recent one or the oldest one. We directly use their implementation to generate baseline results. Let us choose the test distribution at time point $t$. We use all the previous time points $(1,t-1)$ for training. For each source package, we fix a correlation probability~\cite{seqOfsets:2018} and perform Monte-Carlo simulation runs to predict a developer in each run. We perform 20 such runs and prepare a ranked list based on the number of occurrences of a developer across these runs (the larger the number of occurrences of a developer across these runs the better is her rank). We perform this experiment for correlation probabilities in the range $[0.1,0.9]$ in steps of 0.1. We report the results for that correlation probability where the MRR obtained is maximum. 

\noindent{\bf {\em Majority}}: For each source package, we rank the developers based on the number of times they feature in the last $K$ ($K=1, 5$, all) distributions (the results are reported for $K=1$ which turned out to be the best among all choices).  In the autoregressive case, for each source package, a developer present the highest number of times in last $K$ distributions receives better rank and so on. In case of the autoregressive + dependency approach, for each source package, we extend our candidate developer set with the developers of its in(out)-neighbors in previous $K$ distributions. Further, we rank the developers of this set based on the number of times they worked on the target source package in last $K$ distributions. Once again, we use the MRR metric to evaluate this approach.

\noindent{\bf {\em Experimental setup for our method}}: We use Algorithm~\ref{algo:algo1} to rank the candidate developers using autoregressive and autoregressive + network dependency features. For both the models (i.e., LR and MLP), we try different values of parameters. Through grid search we set the number of epochs to 10 and the learning rate to 0.005. The batch size in our experiment is set to 1. The initial values of $\alpha$ and $\beta$ are 1 and 0 respectively\footnote{We also tried other values of $\alpha$ and $\beta$ but they did not affect the results.}. We present the results\footnote{Changes in the value of $K$ does not affect the final results.} for $K=5$. For paucity of space we only report the results for the best combination of features and models; in specific, the LR model with autoregressive features and the MLP model with autoregressive + dependency features.

\noindent{\bf {\em Evaluation}}: The main results are noted in Table~\ref{tab:WeVsSeqOfSets}. We observe that our methods outperform both the majority and the sequence of sets baseline and are closest to the upper bound. Further, the inclusion of network features always brings additional benefits. For all the three distributions, the results from our model (autoregressive) are better from (a) the sequence of sets baseline ($p<0.001$, Mann-Whitney U test) and (b) the majority baseline ($p<0.05$, Mann-Whitney U test). Further, for `Wily' and `Yakkety', the results from our model (autoregressive + dependency) are better than the majority (+ dependency) baseline ($p<0.01$, Mann-Whitney U test).

\section{Discussion and conclusion}
\label{conc}
In this paper we introduced a novel dataset of Ubuntu distributions, motivated by two important software engineering problems: (a)~predicting the urgency of a bug and (b)~recommending a suitable developer for a package. For both the problems we identify a set of simple autoregressive features which themselves are found to be performing very well. Augmenting these features with the dependency network features brings additional benefits.
In future, we would like to investigate further into the dataset to identify if patterns of special relationships exist between developers and bugs and how do these change over time. Discovery of such patterns might allow us to solve the two problems jointly and study other comparable data sets.
\section{Acknowledgement}
Soumen Chakrabarti acknowledges support from a Jagadish Bose Fellowship and a Halepete Family Chair.  Animesh Mukherjee acknowledges a Humboldt Fellowship and the A K Singh Chair.  Pawan Goyal acknowledges support from a Google India AI/ML Research Award. 
%
%
%
\bibliographystyle{splncs04}
\bibliography{ref}
\end{document}